\documentclass[prl,aps,twocolumn,preprintnumbers,amsmath,amssymb]{revtex4}
\usepackage{graphicx}
\usepackage{epsf}
\begin{document}
%
%
\def\be{\begin{equation}}
\def\l{\left}
\def\r{\right}
\def\la{\langle}
\def\ra{\rangle}
\def\ee{\end{equation}}
\def\bc{\begin{center}}
\def\ec{\end{center}}
\def\bea{\begin{eqnarray}}
\def\eea{\end{eqnarray}}
\def\dd{\displaystyle}
\def\nn{\nonumber}
\def\ov{\overline}
\def\cP{{\cal P}}
\def\G{{\cal G}}
\def\cL{{\cal L}}
\def\ad{\dot{\alpha}}
\def\ov{\overline}
\def\cS{{\cal S}}
\def\tphi{{\widetilde{\varphi}}}
\def\tg{{\widetilde{g}}}
\def\dy{{\partial_y}}
\def\de{\partial}
\def\ds{\displaystyle}
\preprint{ROMA-1413/05}
\title{de-Sitter vacua via consistent D-terms}
\author{Giovanni Villadoro}
\affiliation{
Dipartimento di Fisica, Universit\`a di Roma 'La Sapienza' and 
\\ 
INFN, Sezione di Roma, P.le Aldo Moro~2, I-00185 Rome, Italy}
%
\author{Fabio Zwirner}
\affiliation{
Dipartimento di Fisica, Universit\`a di Roma 'La Sapienza' and 
\\ 
INFN, Sezione di Roma, P.le Aldo Moro~2, I-00185 Rome, Italy}
%
\date{August 23, 2005}
%
%
\begin{abstract}
We introduce a new mechanism for producing locally stable de-Sitter or
Minkowski vacua, with spontaneously broken $N=1$ supersymmetry and no
massless scalars, applicable to superstring and $M$-theory
compactifications with fluxes.  We illustrate the mechanism with a
simple $N=1$ supergravity model that provides parametric control on
the sign and the size of the vacuum energy. The crucial ingredient is
a gauged $U(1)$ that involves both an axionic shift and an R-symmetry,
and severely constrains the F- and D-term contributions to the
potential.
\end{abstract}
%
\maketitle
%
%
\section{Introduction}

Superstring and $M$-theory vacua with exact or spontaneously broken
$N=1$ supersymmetry deserve special attention. Their effective $D=4$
supergravities admit chiral fermions and inherit some strong symmetry
properties from the underlying higher-dimensional theory. So far,
perturbative compactifications with fluxes and branes could produce,
at best, either Minkowski vacua of the no-scale type, with broken
supersymmetry and at least one complex flat direction, or
anti-de-Sitter (adS) vacua with all geometrical moduli stabilized.

It is important to go further, exploring the possible existence of
locally stable de-Sitter (dS) or Minkowski vacua, with no residual
flat directions. Some interesting attempts along these lines do indeed
exist \cite{kklt, bkq, dudas, stv}. Most of them rely on the positive
contributions to the potential associated with the gauge symmetry of
the theory, the so-called D-terms. However, the subtle consistency
requirements dictated by the coexistence of the two local symmetries,
supersymmetry and the gauge symmetry, are known on general grounds
\cite{Dbooks, Dpapers}, but were never thoroughly examined in this
context.

Ref.~\cite{kklt} used a superpotential motivated by non-perturbative
effects to produce a supersymmetric adS vacuum, then uplifted the
vacuum energy by a positive contribution to the potential ascribed to
$\ov{D3}$ branes. So far, however, no consistent supergravity
description of such a mechanism was found.  Ref.~\cite{bkq} proposed
to replace the $\ov{D3}$-brane contribution with a D-term contribution
originated by magnetic fluxes but, as will be clear in the following,
such proposal does not fulfill the above-mentioned consistency
requirements. Ref.~\cite{dudas} introduced non-perturbative
superpotentials and D-terms that satisfy all known consistency
conditions, but did not perform a full minimization of the
supergravity potential with respect to all fields.

In the present letter, we first review the general consistency
conditions associated with D-terms in $N=1$ supergravity. We then
construct a simple explicit model fulfilling all such conditions. We
show that, for a wide range of parameters, the model admits locally
stable vacua with spontaneously broken supersymmetry and positive or
negative vacuum energy. The vacuum energy can be zero, or very small
and positive, for special values of the parameters. The key features
of the model can be present in flux compactifications of superstring
theories. The crucial one is a gauged $U(1)$ symmetry that combines an
R-symmetry with an axionic shift, and severely constrains the form of
the F- and D-term contributions to the potential. The rigid version of
such a symmetry, and some of its consequences if unbroken, were
previously studied in \cite{porkou}. Supergravity models with gauged
R-symmetry \cite{dzf} were considered in \cite{gaugedr} \cite{Dpapers}. 
The models of \cite{dudas} have instead a gauged axionic symmetry but no
gauged R-symmetry. We finally comment on the extension of our
mechanism to more general models and on its string/$M$-theory
embedding, leaving a detailed exploration for future work.

\section{D-terms in $N=1$ supergravity}

The gauge-invariant two-derivative action for $N=1$, $D=4$
supergravity with chiral multiplets $\phi^i \sim (z^i, \psi^i)$ and
vector multiplets $V^a \sim (\lambda^a, A_\mu^a)$ is completely fixed
by three ingredients \cite{Dbooks}. The first is the real and
gauge-invariant K\"ahler function $G$, which can be written in terms
of a real K\"ahler potential $K$ and a holomorphic superpotential $W$
as
\be 
G = K + \log |W|^2 \, .  
\label{ggen}
\ee
The second is the holomorphic gauge kinetic function $f_{ab}$, which
transforms as a symmetric product of adjoint representations, plus a
possible imaginary shift associated with anomaly
cancellation. Generalized Chern-Simons terms may also be needed
\cite{afl}, but they will not play any r\^ole in the simple case
discussed in this paper. The third are the holomorphic Killing vectors
$X_a = X_a^i (z) (\de/\de z^i)$, which generate the analytic
isometries of the K\"ahler manifold for the scalar fields that are
gauged by the vector fields. In the following it will suffice to think
of $G$, $f_{ab}$ and $X_a$ as functions of the complex scalars $z^i$
rather than the superfields $\phi^i$.

The gauge transformation laws and covariant derivatives for the
scalars in the chiral multiplets read
\be
\delta z^i  =  X_a^i \, \epsilon^a \, , 
\qquad
D_\mu z^i = \de_\mu z^i - A^a_\mu X_a^i \, ,
\ee
where $\epsilon^a$ are real parameters. The scalar potential is
\be 
\label{vgen}
V = V_F + V_D = e^G \l (G^i G_i - 3 \r) + \frac12 D_a D^a \, ,
\ee
where $G_i = \de G / \de z^i$, scalar field indices are raised with
the inverse K\"ahler metric $G^{i \ov{k}}$, gauge indices are raised
with $[(Re f)^{-1}]^{ab}$, and $D_a$ are the Killing potentials, real
solutions of the complex Killing equations:
\be 
X_a^i = - i \, G^{i \ov{k}} \, 
\frac{\de D_a}{\de \ov{z}^{\ov{k}}} \, .
\ee
The general solution to the Killing equation for $D_a$, compatible
with gauge invariance, is then
\be
\label{eq:solD} 
D_a = i \, G_i \, X_a^i = i \, K_i \, X_a^i + 
i \, \frac{W_i}{W} \, X_a^i \, .  
\ee

Gauge invariance of $G$ requires that $K$ and $W$ be invariant up to a
K\"ahler transformation
\be
K' = K + H + \ov H \, ,
\qquad
W'  =  W \, e^{-H} \, ,
\ee
where $H$ is a holomorphic function, thus it will not be restrictive
to assume that $K$ is gauge invariant. If $W$ is also gauge-invariant,
Eq.~(\ref{eq:solD}) reduces to the standard form
\be 
D_a = i \, K_i \, X_a^i \, .
\ee
Otherwise, it must be
\be
i \, \frac{W_i}{W} \, X_a^i = \xi_a \, ,
\qquad
(\xi_a \in \mathbb{R}) \, ,
\ee
so that the gauge non-invariance of $W$ can be at most an overall
phase with real parameter $\xi_a$, for the Abelian factors
$U(1)_a$. The constants $\xi_a$ correspond to gaugings of the
R-symmetry, and give rise to the supergravity expression for the
D-terms \cite{Dbooks, Dpapers}:
\be 
D_a = i \, K_i \, X_a^i + \xi_a \, .
\ee
The $\xi_a$ are then the genuine Fayet-Iliopoulos (FI) terms of
supergravity. For a linearly realized gauge symmetry, $i \, K_i \,
X_a^i = - K_i \, (T_a)^i_{\; k} z^k$, and we recover the standard
expression of \cite{cremmer} for the D-terms. For an axionic
realization, $X_a^i = i \, q_a^i$, where $q_a^i$ is a real constant,
and we obtain the so-called field-dependent FI terms.

A few comments are now in order. Eq.~(\ref{eq:solD}) shows that
D-terms are actually proportional to F~terms, $F_i = e^{G/2} \,
G_i$. Two facts, frequently forgotten in the recent literature,
become then obvious. First, and in contrast with the rigid case, there
cannot be pure D-breaking of supergravity, unless the gravitino mass
vanishes and the D-term contribution to the vacuum energy is
uncanceled, as in the limit of global supersymmetry. Second, if $V_F$
admits a supersymmetric adS vacuum configuration, $\la G_i \ra = 0$
($\forall i$) and $\la e^G \ra \neq 0$, such configuration
automatically minimizes $V_D$ at zero. For this kind of vacua, as
already stressed in \cite{dealw}, D-terms cannot be used to raise the
vacuum energy from negative to positive or zero. Moreover, for the
theory to be consistent, $W$ must be gauge-invariant, up to an overall
phase for $U(1)$ factors. This severely restricts the possibility of
constructing superstring-inspired supergravity models with both
non-perturbative superpotentials and D-terms. D-terms associated with
a gauged $U(1)$ symmetry cannot coexist with `racetrack' (sums of
exponentials) or other (e.g. constant plus exponential)
superpotentials, when the latter break such a symmetry. This is in
agreement with some recent results in superstring compactifications,
where it was shown that the gauged isometries are protected from being
broken, both by instanton-induced \cite{kpt} and by flux-induced
\cite{cfi} superpotentials. On the other hand, a rigid axionic and/or
R-symmetry, possibly the remnant of a gauged symmetry spontaneously
broken at the string scale, can be explicitly broken to a discrete
subgroup by non-perturbative effects.

\section{A model with stable de-Sitter vacua}

Consider a model with a single chiral multiplet ${\cal S} \sim (S ,
\chi)$ and K\"ahler potential
\be
\label{kamod}
K = - p \,  \log \l( S + \ov{S} \r) + K_0 \, ,
\qquad (0 < p \in {\mathbb R}) \, ,
\ee
with $K_0$ real and $S$-independent. This form of $K$ is familiar from
superstring compactifications. Decomposing the complex scalar as $S=s
+ i \, \sigma$, $s$ can stand here for the string dilaton, a
geometrical (K\"ahler or complex structure) modulus of the
compactification manifold, or a combination thereof; $\sigma$ could
instead carry the degree of freedom of some component of the NS-NS or
R-R potentials, or even of the internal metric. The $U(1)$ isometry
acting as a shift on the `axion' $\sigma$ can be gauged by a vector
multiplet. The corresponding holomorphic Killing vector is just an
imaginary constant,
\be
\label{kimod}
X^S = i \, q \, ,
\qquad
(q \in {\mathbb R}) \, .
\ee
The most general form of the superpotential compatible with the gauged
$U(1)$ symmetry is then
\be
\label{wmod}
W = W_0 \, e^{- k \, S} \, ,
\qquad
(k \in {\mathbb R}) \, ,
\ee
where $W_0$ is $S$-independent. Eq.~(\ref{wmod}) has the typical form
of the non-perturbative superpotentials induced by instantons or
gaugino condensation \cite{nonpertw}. Notice that the gauged $U(1)$ is
a combination of the axionic $U(1)$, acting as a shift on $\sigma$ and
leaving all the other fields invariant, and the $U(1)$ R-symmetry,
with charge $+(\xi/2)$ for $\chi$, $-(\xi/2)$ for the gravitino
$\psi_\mu$ and the gaugino $\lambda$, and zero for all the bosonic
fields. $\xi = k \, q$ is the constant FI term. For the gauge kinetic
function, we take
\be
\label{fmod}
f = S \, ,
\ee
as typical of superstring compactifications, but the model would still
work for $f = a \, S + b$, the most general form compatible with the
gauged symmetry. Notice, finally, that the gauge kinetic function $f$ 
of the possible gauge group factor associated with instantons
or gaugino condensation does not need to coincide
with the $U(1)$ gauge kinetic function of eq.~(13):
this may be of help in obtaining vacua at weak
coupling.

The scalar potential of Eq.~(\ref{vgen}) reads then
\be
\label{vfmod}
V_F = \frac{e^{G_0} \, e^{ - 2 \, k \, s}}{ 
\dd (2 s)^p} \, \l[ \frac{\dd (2 s)^2 }{ \dd p} \,
\l( k + \frac{\dd p }{ \dd 2 \, s} \r)^2 - 3 \r] \, , 
\ee
\be
\label{vdmod}
V_D = \frac{\dd q^2}{ \dd 2 s} \,
\l( k + \frac{\dd p }{ \dd 2 \, s} \r)^2 \, ,
\ee
where $e^{G_0} \equiv |W_0|^2 \, e^{K_0}$. As required by gauge
invariance, $V$ does not depend on $\sigma$: the axion is absorbed by
the massive $U(1)$ vector boson via the Higgs effect
\cite{dsw}. However, both $V_F$ and $V_D$ depend non-trivially on
$s$. For $k < 0$, there is always a stable supersymmetric adS vacuum
at $\langle s \rangle = - p /(2 k)$, but there can be also metastable
dS vacua for suitable values of the parameters. For $k>0$ and $p \ge
3$, $V$ is positive definite and monotonically decreasing. For $k>0$
and $p<3$, $V_F$ is unbounded from below for $s \to 0$, but $V_D$ is
positive definite and diverges for $s \to 0$. As a result, for a wide
range of parameters there is a locally stable dS (or stable adS)
minimum of $V$ at a finite value $\langle s \rangle$, with
spontaneously broken supersymmetry. At this level, having approximate
Minkowski vacua requires a tuning of the parameters so that $\langle
V_D \rangle \simeq - \langle V_F \rangle \gg \langle V \rangle$. In
principle, this may find an explanation in the correlations of the
underlying string theory. Notice that $\langle s \rangle$ can be
continuously varied by rescaling the parameters as:
\be
\label{scaling}
k \to \alpha^{-1} \, k \, ,
\quad
e^{G_0} \to \alpha^p \, e^{G_0} \, ,
\quad
q \to \alpha^{3/2} \, q \, ,
\ee
($0 < \alpha \in {\mathbb{R}}$), which leads to $\langle s \rangle \to
\alpha \, \langle s \rangle$. A representative example is shown in
Fig.~1.
\begin{figure}
\flushleft{\includegraphics*[width=3.in]{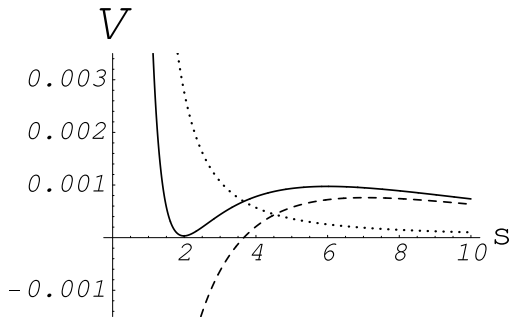}}
\caption{$V$ (solid line), $V_F$ (dashed line) and $V_D$ (dotted line) 
for $p=1$, $q=0.3$, $e^{G_0}=1/64$ and $k =0.1$.}
\label{dsmin}
\end{figure}
\section{Discussion}

The simple model discussed above can be easily generalized. The
inclusion of additional gauge multiplets is straightforward (apart
from anomaly cancellation, see below), thus we consider the inclusion
of additional chiral multiplets $\phi^i$, transforming linearly under
the axionic $U(1)$, with charges $q^i$. Since the R-charge is fixed to
be vanishing for the $z^i$ and $+(\xi/2)$ for the $\psi^i$, the
corresponding charges under the gauged $U(1)$ will be $q^i$ and $(q^i
+ \xi/2)$, respectively.

Consider first the simple case where the full $K$ can be written as in
(\ref{kamod}), with $K_0$ real gauge-invariant function of the $z^i$.
Assume also a factorized $W$ as in (\ref{wmod}), with $W_0$ analytic
gauge-invariant function of the $z^i$, and $f$ as in (\ref{fmod}). We
may think of the $z^i$ as other moduli of string/$M$-theory
compactifications, orthogonal to $S$, or the
scalar fields of the Minimal Supersymmetric Standard Model (MSSM).
It is easy to see that, under mild and plausible assumptions, the
minimization of the potential $V$ with respect to the $z^i$ and $S$
can be fully decoupled. If there is a field configuration $\langle z^i
\rangle$ such that $\langle G_i \rangle = 0$ and $ q^i \, \langle z^i
\rangle = 0$ $\forall i$, then $\langle s \rangle$ as determined in
the previous section and $\langle z^i \rangle$ extremize the full
potential $V$, with no mass mixing between the $z^i$ and $s$. A
locally stable minimum can be obtained for a wide range of the
parameters in $G_0$, with the same $\langle V \rangle$ as before. If,
instead, there is a minimum of $V$ such that $\langle V_D \rangle =
0$, then we can decouple a massive $N=1$ vector multiplet and discuss
a simpler but less interesting effective theory.

The previous model can be further generalized by including additional
chiral multiplets ${\cal C}^\alpha \sim (C^\alpha , \psi^\alpha)$,
which could stand for some or all the MSSM squarks and leptons, in the
approximation of small field fluctuations about $\langle C^\alpha
\rangle = 0$, but relaxing the factorization of $K$ and $W$. For
example, we could add to $K$ a $\Delta K = \sum_\alpha |C^\alpha|^2 (S
+ \ov{S})^{n_\alpha}$ ($n_\alpha \in \mathbb{Z}$), and to $W$ a
$\Delta W$ polynomial in the $C^\alpha$ and transforming with the same
phase as $e^{- k \, S}$, e.g. $\Delta W = d_{\alpha \beta \gamma}
C^\alpha C^\beta C^\gamma$ with $q^\alpha + q^\beta + q^\gamma = - k
\, q$. Also in this case, for suitable values of the new parameters,
there is a local minimum of the full potential $V$ with $\langle
C^\alpha \rangle = \langle G_\alpha \rangle = 0$ $\forall \alpha$,
positive squared masses for all the new scalar fields $C^\alpha$, and
all the remaining features as in the previous model.

For a consistent effective theory, all gauge and gravitational
anomalies associated with our gauged $U(1)$ must vanish: in
particular, the cubic (${\cal A}_{U(1)^3}$), the gravitational (${\cal
A}_{U(1)}$) and the mixed-gauge anomaly (${\cal A}_{U(1)\,\G^2}$) if
the full gauge group is $U(1) \times \G$. The fermionic contributions
to the cubic and gravitational anomalies are:
\bea
Tr\,Q^3 & = & 3\l(-\frac{\xi}{2}\r)^3
+\sum_{i,\alpha} \l(q^{i,\alpha} +
\frac{\xi}{2}\r)^3 \,, 
\\ 
Tr \, Q & = & - 21 \l( -\frac{\xi}{2}\r) 
+\sum_{i,\alpha} \l(q^{i,\alpha} +\frac{\xi}{2}\r) \,, 
\eea
where the contributions from $\lambda$ and $\chi$ cancel each other
and have been omitted. The remaining ones come from $\psi_\mu$ (see
\cite{anom32}) and possible $\psi^{i, \alpha}$, respectively. These
contributions must cancel the Green-Schwarz (GS) contributions
\cite{gs} coming from the variation of $\sigma$ and proportional to
$q$. All the resulting conditions are model dependent, in particular:
all of them depend on the matter content; the GS contribution to
${\cal A}_{U(1)}$ depends on higher derivative terms ($R^2$); ${\cal
A}_{U(1)\, \G^2}$ depends also on the details of $\G$. However, there
are in principle strong combined constraints on the possible matter
content and on the parameters $k$ and $q$.

The mechanism discussed in this letter should be relevant for the study
of superstring and M-theory vacua, with the anomaly constraints
automatically satisfied and the possibility of determining $k$ and
$q$. $N=1$ supergravities obtained from compactifications with fluxes
generically allow for some shift symmetry to be gauged. 

In the heterotic theory, the shift symmetry of the universal axion
\cite{wittens}, dual to $B_{\mu \nu}$, is gauged via the GS mechanism,
and fluxes can be used to generate a superpotential $W_0$
\cite{glm,dkpz}
\be
\int_{X_6} (H + i \, dJ ) \wedge \Omega \, ,
\ee
that can stabilize all geometrical moduli with vanishing F-terms and
positive masses. A modification of $W$ (or, equivalently, of $K$) as
in (\ref{wmod}) would then stabilize also the dilaton $S$ on a dS
vacuum, breaking the local symmetries only spontaneously.

Also in type-IIA compactifications with fluxes, the superpotential
\cite{vz} \cite{dkpz}
\be
\int_{X_6} \mathbf{G} \, e^{iJ}-i (H-idJ) \wedge \Omega \, ,
\ee
can produce the stabilization of all bulk moduli with vanishing
F-terms, with the exception of at least one massless axion \cite{vz}
\cite{cfi}, associated with a shift symmetry that can eventually be
gauged. In this case the role of $S$ is played by a linear combination
of the dilaton and the complex structure moduli $\Omega$, and its
dependence cannot be factorized from the other moduli anymore. Whether
in this case an analogous modification of the superpotential would
allow the lifting of the vacuum energy and the stabilization of all
moduli remains an open problem.

Finally, it would be interesting to understand better how the needed
superpotential modifications actually originate from string/M-theory,
and what the corresponding constraints on the various parameters are.

\section{Acknowledgments}
We thank S.~Ferrara for discussions and M.~Porrati for comments. This
work was supported in part by the European Programme ``The Quest For
Unification'', contract MRTN-CT-2004-503369.
%
%

%
\end{document}